%% file: main.tex
\newif\ifiacr
\newif\ifarxiv
\newif\ifacm

\arxivtrue
\acmfalse
\iacrfalse

\ifiacr
\documentclass{llncs}
\usepackage{caption}
\usepackage{subcaption}
\usepackage{cite}
\else
\documentclass[sigconf]{acmart}
\usepackage{caption, subcaption}
\fi

\def\BibTeX{{\rm B\kern-.05em{\sc i\kern-.025em b}\kern-.08emT\kern-.1667em\lower.7ex\hbox{E}\kern-.125emX}}

\usepackage{microtype}
\usepackage{graphicx}
\usepackage{makecell}
\usepackage{booktabs} % for professional tables
\usepackage{multirow}
\usepackage{listings}
\usepackage{graphicx}
\usepackage{color}

\usepackage{hyperref}

\usepackage{pifont}% http://ctan.org/pkg/pifont
\newcommand{\cmark}{\ding{51}} % Checkmark
\newcommand{\xmark}{\ding{55}} % X (complement to checkmark)
\usepackage{stfloats} % For two-column image in appendix
\usepackage{soul} % For highlighting
\usepackage{tikz}
\usepackage[normalem]{ulem} % underline
\usepackage[separate-uncertainty=true]{siunitx} % To align table by \pm
\usepackage{siunitx}
\usepackage{amsmath}
\usepackage{amsfonts}
\usepackage{amssymb}
\usepackage[nointegrals]{wasysym} % For wide symbols
\usepackage{booktabs}

\definecolor{mygreen}{rgb}{0,0.6,0}
\definecolor{mygray}{rgb}{0.5,0.5,0.5}
\definecolor{mymauve}{rgb}{0.58,0,0.82}

\ifarxiv
\renewcommand\footnotetextcopyrightpermission[1]{} % removes footnote with conference information in first column
\fi

\ifiacr
\fi

\ifacm
	\copyrightyear{2019}
	\acmYear{2019}
	\setcopyright{acmcopyright}
	\acmConference[CF '19]{Proceedings of the 16th conference on Computing Frontiers}{April 30-May 2, 2019}{Alghero, Italy}
	\acmBooktitle{Proceedings of the 16th conference on Computing Frontiers (CF '19), April 30-May 2, 2019, Alghero, Italy}
	\acmPrice{15.00}
	\acmDOI{10.1145/3310273.3323047}
	\acmISBN{978-1-4503-6685-4/19/05}
\fi

\begin{document}

\ifiacr
\title{nGraph-HE: A Graph Compiler for Deep Learning on Homomorphically Encrypted Data}

\author{Fabian Boemer\inst{1}  \and
	Yixing Lao\inst{1}  \and
	Rosario Cammarota\inst{1} \and
	Casimir Wierzynski\inst{1}}

\institute{
	Intel AI Research\\
	San Diego, CA \\
	\email{
		fabian.boemer@intel.com}
	}

\maketitle

\else

% The "title" command has an optional parameter, allowing the author to define a "short title" to be used in page headers.
\title[nGraph-HE]{nGraph-HE: A Graph Compiler for Deep Learning on Homomorphically Encrypted Data}

\author{Fabian Boemer}
\email{fabian.boemer@intel.com}
\affiliation{%
\institution{Intel AI Research}
\city{San Diego}
\state{California}
\country{USA}
}

\author{Yixing Lao}
\email{yixing.lao@intel.com}
\affiliation{%
\institution{Intel AI Research}
\city{San Diego}
\state{California}
\country{USA}
}

\author{Rosario Cammarota}
\email{rosario.cammarota@intel.com}
\affiliation{%
\institution{Intel AI Research}
\city{San Diego}
\state{California}
\country{USA}
}

\author{Casimir Wierzynski}
\email{casimir.wierzynski@intel.com}
\affiliation{%
\institution{Intel AI Research}
\city{San Diego}
\state{California}
\country{USA}
}
\fi

% Blind footnote
\newcommand\blfootnote[1]{%
	\begingroup
	\renewcommand\thefootnote{}\footnote{#1}%
	\addtocounter{footnote}{-1}%
	\endgroup
}

%
% By default, the full list of authors will be used in the page headers. Often, this list is too long, and will overlap
% other information printed in the page headers. This command allows the author to define a more concise list
% of authors' names for this purpose.
%\renewcommand{\shortauthors}{F. Boemer et al.}
%
% The abstract is a short summary of the work to be presented in the article.
\begin{abstract}
\input{abstract}
\ifarxiv
 \blfootnote{To appear in ACM International Conference on Computing Frontiers 2019.}
 \fi
 \ifiacr % IACR uses hyper-ref, so blind footnote looks bad
 \footnote{To appear in ACM International Conference on Computing Frontiers 2019.}
 \fi
\end{abstract}

% Indexing for ACM categories
% Generated at http://dl.acm.org/ccs.cfm
\ifiacr
\else

\fi

% presented. Separate the keywords with commas.

%
% This command processes the author and affiliation and title information and builds
% the first part of the formatted document.
\ifiacr
\else
\maketitle
\fi

\input{body}

\clearpage
\ifiacr
\bibliographystyle{splncs04}
\else
\bibliographystyle{ACM-Reference-Format}
\fi
\bibliography{main}

% If your work has an appendix, this is the place to put it.
%\clearpage
\appendix
\include{appendix}

\clearpage
\include{artifacts}

\end{document}

%% file: abstract.tex
Homomorphic encryption (HE)---the ability to perform computation on encrypted data---is an attractive remedy to increasing concerns about data privacy in deep learning (DL). However, building DL models that operate on ciphertext is currently labor-intensive and requires simultaneous expertise in DL, cryptography, and software engineering. DL frameworks and recent advances in graph compilers have greatly accelerated the training and deployment of DL models to various computing platforms. We introduce nGraph-HE, an extension of nGraph, Intel's DL graph compiler, which enables deployment of trained models with popular frameworks such as TensorFlow while simply treating HE as another hardware target. Our graph-compiler approach enables HE-aware optimizations-- implemented at compile-time, such as constant folding and HE-SIMD packing, and at run-time, such as special value plaintext bypass. Furthermore, nGraph-HE integrates with DL frameworks such as TensorFlow, enabling data scientists to benchmark DL models with minimal overhead.

%% file: body.tex
\input{introduction}

\input{implementation}

\input{results}

\input{extensions}

%\section*{Acknowledgements}
%
%\textbf{Do not} include acknowledgements in the initial version of
% the paper submitted for blind review.
%
% If a paper is accepted, the final camera-ready version can (and
% probably should) include acknowledgements. In this case, please
% place such acknowledgements in an unnumbered section at the
% end of the paper. Typically, this will include thanks to reviewers
% who gave useful comments, to colleagues who contributed to the ideas,
% and to funding agencies and corporate sponsors that provided financial
% support.

%% file: introduction.tex
\section{Introduction}

One of the key challenges in deploying machine learning (ML) at scale is how to help data owners learn from their data while protecting their privacy. This issue has become more pressing with the advent of regulations such as the General Data Protection Regulation~\cite{voigt2017eu}. It might seem as though “privacy-preserving machine learning” would be a self-contradiction: ML wants data, while privacy hides data~\cite{wierzynski2018fortune}. One promising solution to this problem is known as homomorphic encryption (HE). Using HE, one can perform computation on encrypted data without decrypting it. Data owners can encrypt their data with the public key, send it to a data processor that has no access to the secret key, and receive the answer to their query in encrypted form, which only the data owner can unlock with the secret key.

The idea of HE dates back to 1978~\cite{rivest1978data}, and theoretical breakthroughs occurred in 2009~\cite{gentry2009fully} to make the idea real but highly impractical. Further algorithmic breakthroughs have occurred since then, in tandem with the development of post-quantum cryptosystems and their implementations~\cite{bernstein2017post, DBLP_journals_csur_NejatollahiDRRB19} to yield HE schemes that map naturally onto vector addition and multiplication---the core of DL workloads. Recent work has shown the feasibility of evaluating convolutional neural networks using lattice-based HE cryptosystems~\cite{gilad2016cryptonets, minionn,hesamifard2017cryptodl,juvekar2018gazelle, sanyal2018tapas, chou2018faster}.

One of the biggest accelerators in DL has been the development and rapid adoption of software frameworks, such as TensorFlow~\cite{abadi2016tensorflow}, MXNet~\cite{chen2015mxnet} and PyTorch~\cite{pytorch}, making use of open-source graph compilers such as Intel nGraph~\cite{cyphers2018intel}, XLA~\cite{xla} and TVM~\cite{tvm}, that allow data scientists to describe DL networks and operations at a high level while hiding details of their software and hardware implementation. By contrast, a key challenge for building privacy-preserving DL systems using HE has been the lack of such a framework. As a result, developing and deploying DL models that operate on ciphertext is currently labor intensive and forces data scientists to become experts in DL, cryptography, and software engineering.

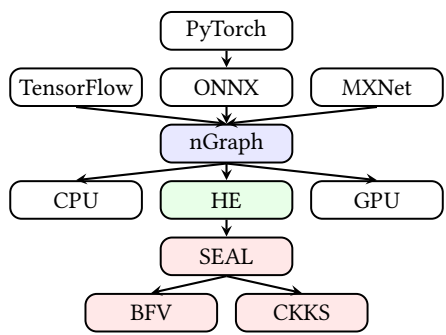
\begin{figure}
 \begin{center}
\tikzset{every picture/.style={line width=0.75pt}} %set default line width to 0.75pt

\tikzset{
	arrow/.style = {
		thick,->,>=stealth
	},
	doublearrow/.style ={
		thick, <->, >=stealth
	},
	arrowline/.style = {
		thick
	},
	object/.style ={
		rectangle, minimum width=1.75cm, minimum height=0.5cm, text centered, draw=black,rounded corners=.15cm
	},
	ng_object/.style ={
		rectangle, minimum width=1.75cm, minimum height=0.5cm, text centered, draw=black,rounded corners=.15cm, fill={rgb:blue,1;white,10}
	},
	he_object/.style ={
		rectangle, minimum width=1.75cm, minimum height=0.5cm, text centered, draw=black,rounded corners=.15cm, fill={rgb:green,1;white,10}
	},
	seal_object/.style ={
		rectangle, minimum width=1.75cm, minimum height=0.5cm, text centered, draw=black,rounded corners=.15cm, fill={rgb:red,1;white,10}
	}
}
\centering

\begin{tikzpicture}[y=0.75cm]
\node(bfv) at (1,0) [seal_object]{BFV};
\node(ckks) at (3,0) [seal_object]{CKKS};
\node(seal) at (2, 1) [seal_object]{SEAL};
\node(he) at (2, 2) [he_object]{HE};
\node(cpu) at (0, 2) [object]{CPU};
\node(gpu) at (4, 2) [object]{GPU};
\node(ngraph) at (2, 3) [ng_object]{nGraph};
\node (tensorflow) at (0, 4) [object]{TensorFlow};
\node (onnx) at (2, 4) [object]{ONNX};
\node (mxnet) at (4,4) [object]{MXNet};
\node (pytorch) at (2,5) [object]{PyTorch};

\draw [arrow] (pytorch.south) -- (onnx.north);
\draw [arrow] (onnx.south) -- (ngraph.north);
\draw [arrow] (mxnet.south) -- (ngraph.north);
\draw [arrow] (tensorflow.south) -- (ngraph.north);
\draw [arrow] (ngraph.south) -- (cpu.north);
\draw [arrow] (ngraph.south) -- (gpu.north);
\draw [arrow] (ngraph.south) -- (he.north);
\draw [arrow] (he.south) -- (seal.north);
\draw [arrow] (seal.south) -- (ckks.north);
\draw [arrow] (seal.south) -- (bfv.north);
\end{tikzpicture}
\caption{Overview of the nGraph-HE software stack. nGraph-HE currently supports the SEAL encryption library~\cite{seal}, and the underlying cryptosystems BFV~\cite{bfv} and CKKS~\cite{ckks}, and it can be extended to support additional cryptosystems.}
\label{fig:ngraph}
\end{center}
\end{figure}

In this work, we leverage recent work in graph compilers to overcome this challenge. Specifically, we present nGraph-HE, an HE backend to the Intel nGraph DL graph compiler that allows data scientists to train networks on the hardware of their choice in plaintext, then easily deploy these models to HE cryptosystems that operate on encrypted data. The core idea is to create a privacy-preserving hardware abstraction layer, with its own instruction set architecture (ISA) (Section~\ref{sec:assembly}) and optimization support (Section~\ref{sec:optim}). This hides the complexity of HE from data scientists while exploiting the considerable compiler tooling and DL frameworks that the DL community has built (Figure~\ref{fig:ngraph}). Using this approach, for example, modifying an existing TensorFlow model to operate on encrypted data becomes as easy as adding a single line of code (Appendix~\ref{sec:appendix}). Indeed, the open-source release of this framework\footnote{The nGraph-HE library is available under the Apache 2.0 license at https://ngra.ph/he}
	has already gathered significant attention in the DL community~\cite{venturebeat, dropoutlabs}.

Using HE to implement DL computations imposes a number of constraints due to the mathematical requirements of HE and DL themselves, such as limited arithmetic depth and polynomial activation functions (Section~\ref{sec:challenges}). Overcoming these constraints is an area of active algorithmic research~\cite{gilad2016cryptonets, minionn,hesamifard2017cryptodl, juvekar2018gazelle, sanyal2018tapas, chou2018faster}. The contributions of this paper are along a different vector, namely, how to provide a software framework for developing privacy-preserving DL models that cleanly separates DL and HE functions (Figure~\ref{fig:ngraph}). This will enable the DL and HE communities to improve their own technologies as independently as possible while still enjoying the advances of the other with minimal changes to the high-level code.

In this paper, we present the following:
\begin{enumerate}
\item We describe an efficient software framework for combining DL and HE. To our knowledge, we present the first use of a DL graph compiler and intermediate representation (IR) to accelerate the development and deployment of privacy-preserving machine learning models.
 \item We develop HE-aware graph-compiler optimizations, both at compile-time and at run-time. The compile-time optimizations include graph-level optimizations such as batch-norm folding and parallel operations through HE-SIMD packing and OpenMP parallelization. Runtime optimizations include special plaintext value bypass and ciphertext-plaintext operations.
 
\item We demonstrate the framework on: (1) subgraphs of DL models: general matrix-matrix multiplication (GEMM) operations, and a convolution-batch-norm operation; and (2) two convolutional neural network benchmark problems (MNIST and CIFAR-10) with different choices of encryption parameters, using Python and TensorFlow. Furthermore, we verify that the runtime overhead imposed by the additional software layers is small ($0.1\%$ of total runtime) compared to implementing these operations in C++ using HE libraries directly.
\end{enumerate}

\section{Background}

\subsection{Homomorphic encryption}
\label{sec:he_background}

What does it mean for a cryptosystem to be {\em homomorphic}? Informally, an encryption function $E$ and its decryption function $D$ are homomorphic with respect to a class of functions $\mathcal{F}$ if for any function $f \in \mathcal{F}$, we can construct a function $g$ such that $f(x) = D \left( g(E(x)) \right)$ for some set of $x$ that we care about\footnote{We are omitting the public and secret keys that would also be arguments for the encryption and decryption functions.}. That is, for certain cryptosystems and target functions, it is possible to map a desired computation (the function $f$) on plaintext into a specific computation on ciphertext (the function $g$) whose result, when decrypted, matches the desired plaintext result. For a detailed review of HE, we refer the reader to~\cite{acar2018survey}.

\begin{figure}
\includegraphics[width=0.9\columnwidth]{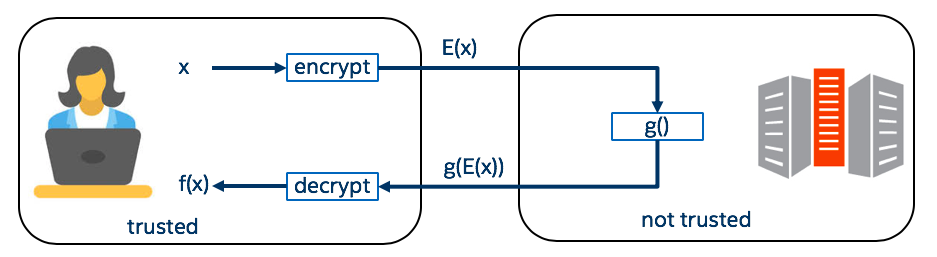}
\caption{Simple model of secure inference via HE.}
\label{fig:cloud}
\end{figure}

 Figure~\ref{fig:cloud} shows how this property enables a user, Alice, to perform inference on private data using a remote, untrusted computer. The remote machine receives a ciphertext from Alice (with no decryption key), executes a function $g$ on the ciphertext, then returns the result to Alice. Alice then decrypts the result to reveal the plaintext for $f(x)$. At no point does the remote machine gain access to Alice's unencrypted data. An analogous setup provides secure inference in the case where the function $g$ is kept private, while the data remains unencrypted, as might occur when, for example, $g$ corresponds to a proprietary 3rd party DL model.

One important property of RLWE-based HE schemes is \emph{semantic security}, which is the inability of a computationally bounded adversary to distinguish between the ciphertexts of known plaintexts. Notably, $E(x) \neq E(y)$, even when $x=y$. This is due to random noise which is introduced during the encryption process. Without this property, a malicious remote server in Figure~\ref{fig:cloud} might be able to deduce $f(x)$ in cases where $g$ maps to a finite number of outputs, as in binary classification problem, by performing inference on inputs whose classification is already known.

\subsection{Challenges of homomorphically encrypted deep learning}
\label{sec:challenges}
HE schemes are often subject to several mathematical limitations:

{\bf Supported functions.} Some HE schemes only support a single algebraic operation, such as addition or multiplication. These are known as ``partially homomorphic'' schemes (PHE). Others schemes, called ``fully homomorphic'' (FHE), support two, such as addition and multiplication. Note that composing addition and multiplication suffices to construct polynomial functions, and hence polynomial approximations to non-polynomial functions such as sigmoid or ReLU\footnote{Going further, by building gates out of addition and multiplication over GF(2), one can in theory implement any boolean circuit, and hence any computable function.}. Notably, this limitation prevents the exact computation of any comparison-based operations, such as Max, Min, and ReLU, as well as common functions such as exponential or sigmoid. One workaround to this limitation in the case of a final softmax layer is to leave the softmax calculation to our user Alice after she decrypts the model outputs. Finally, ``leveled homomorphic'' schemes (LHE) support addition and multiplication, but only up to a fixed computational depth. 

{\bf Computational depth.} HE schemes derived from Gentry's original lattice-based system~\cite{gentry2009fully} rely on noise to hide plaintext. This encryption noise tends to accumulate with each homomorphic operation, and decryption becomes impossible if this noise exceeds a threshold. One common solution to this problem is to constrain the depth of the computation and set encryption parameters accordingly. Other solutions involve noise management techniques such as {\em bootstrapping}, which, depending on the HE scheme, may incur significant computational costs, but can extend the computational depth indefinitely. LHE schemes do not perform bootstrapping, relying instead on the fixed computational depth of DL models.

{\bf Number fields.} Most HE schemes operate over integers~\cite{helib,seal}, while others use booleans~\cite{tfhe} or real numbers~\cite{heaan}. One particular challenge in the case of integer-based schemes is scaling the magnitude of numbers by factors less than 1. Most DL models require real, i.e., non-integer, numbers, so adapting integer HE schemes typically involves mapping large integers to a fixed-point representation using a scaling factor. Preventing the scaling factor from accumulating, however, requires division by the scaling factor after each multiplication, which is not possible in all HE schemes.

{\bf Computational and memory load.} The cryptographic computations required to implement HE typically consume several orders of magnitude more CPU time and memory compared to their plaintext counterparts. These costs have long been the critique of HE. A detailed response to these critiques is out of the scope of this paper, but we note that there have been dramatic improvements in this area---for example, the runtime for homomorphic inference on the seminal CryptoNets MNIST network has been reduced from 297.5s~\cite{gilad2016cryptonets} to 0.03s~\cite{juvekar2018gazelle} in two years (although the latter uses a hybrid scheme; see Section~\ref{sec:extensions}).

From a software engineering perspective, there is additional complexity: there are multiple libraries for HE~\cite{helib, tfhe, heaan, seal, palisade}, based on multiple HE schemes~\cite{sathya2018review}, and with a variety of APIs (with some notable attempts to provide uniformity~\cite{palisade,bourachimera}). This diversity makes it difficult for developers to evaluate the tradeoffs of different schemes in the context of their specific applications. Moreover, the complexity of implementing DL models has led to the development of multiple DL libraries~\cite{jia2014caffe, tokui2015chainer, chen2015mxnet, abadi2016tensorflow, pytorch, paddlepaddle}. Finally, and not surprisingly, no currently-available DL libraries were designed with HE in mind, and vice-versa. As a result, developers of privacy-preserving DL models have been forced either to import DL functions into HE code, or HE functions into DL code, with large code changes required if either of these library choices should change.

Given the computational and memory overhead of HE, we target inference, rather than training.
The inference use case is also particularly relevant from a privacy point of view, given that statistical techniques such as differential privacy do not easily apply to the case of protecting the privacy of the query.

\subsection{Mathematical Background}
We provide a brief introduction to the mathematical objects used in HE. Many HE schemes are based on the assumed hardness of the Ring Learning with Errors (RLWE) problem~\cite{cryptoeprint:2012:230}, which uses polynomials whose coefficients are members of a finite field.~\cite{brakerski2011fully}. In particular, let $N = 2^n$ for some positive integer $n$. $N$ is known as the \emph{polynomial modulus degree}. Let $\mathcal{R} = \mathbb \mathbb{Z}[x] / (x^N + 1)$ be the ring of integer-coefficient polynomials with degree at most $N$. We denote by $\mathcal{R}_a = \mathbb{Z}_a[x] / (x^N + 1)$ the same ring as $\mathcal{R}$, but whose coefficients are integers modulo $a$. The plaintexts and ciphertexts in the HE schemes we consider consist of pairs or lists of polynomials in $\mathcal{R}_t$ and $\mathcal{R}_q$, where $t, q>0$ are known as the \emph{plaintext modulus} and \emph{ciphertext modulus}, respectively. In practice, $t,q$ are often large enough that, for performance reasons, they are decomposed as $q = \prod_{i} q_i$, $t = \prod_i t_i$, where $q_i$ and $t_i$ are known as \emph{ciphertext coefficient moduli} and \emph{plaintext coefficient moduli}, respectively. Typically, multiplying two ciphertexts $c_1, c_2 \in \mathcal{R}_q^{j+1}$ of size $j+1$ results in a ciphertext of larger size, up to $j^2/2$ in the BV scheme~\cite{brakerski2014efficient}, and $2j+1$ in the BFV and CKKS encryption schemes~\cite{seal}. Subsequent operations on these larger ciphertexts become slower; however, an operation called \emph{relinearization} reduces the length of each ciphertext to mitigate this slowdown. Performing relinearization requires a public \emph{relinearization key}.

These RLWE parameters, including the polynomial modulus degree, ciphertext moduli, and plaintext moduli, are chosen to ensure a \emph{security level}, measured in bits, where $\lambda$-bit security indicates ${\sim}2^\lambda$ operations are required to break the encryption~\cite{van2013estimating, lindner2011better}. The choice of $\lambda$ depends on the security needs of the application, with typical values for $\lambda$ being 128 bits, 192 bits, and 256 bits. The runtime requirements to mitigate several attacks on different security levels are detailed in~\cite{HomomorphicEncryptionSecurityStandard}. Additionally, the parameters need to be chosen sufficiently large such that the amplification of the random noise during arithmetic operations does not render the original message unrecoverable. Specifically, each ciphertext (or plaintext) encoded in $\mathcal{R}_q$ (or $\mathcal{R}_t$) is associated with a level $l$ with, $0 \leq l \leq L$, where $L$ is maximum multiplicative depth. The multiplicative depth $L$ is one of the parameters to the HE scheme in addition to the RLWE parameters. The HE scheme allows at most $l$ multiplications on the ciphertext. Multiplication is typically much more expensive than addition, so the multiplicative depth of the desired computation, $L_f$, is an important consideration when computing on HE. Hence, for an assigned computation of a certain multiplicative depth $L \geq L_f$, HE parameters are selected to guarantee, for example, 128-bit security. The choice of parameters presents a trade-off between security to preserve data privacy, and speed of computation.

\subsection{Related Work}
While there has been much previous work detailing algorithmic improvements to HE for DL ~\cite{gilad2016cryptonets, minionn, juvekar2018gazelle, sanyal2018tapas, hesamifard2017cryptodl, chou2018faster}, there have been only a few notable efforts to provide privacy-preserving machine learning software frameworks. PySyft~\cite{ryffel2018generic-pysyft} is a generic framework for distributed, privacy-preserving DL, built on PyTorch, that uses multi-party computation (MPC) for private computation. TF-encrypted~\cite{dahl2018private-tf-encrypted} also enables private machine learning via MPC, and is built on TensorFlow. Both of these systems are tied to a specific DL framework and use MPC, not HE, which assumes a different security model. By contrast, by operating on computational graphs, nGraph-HE enables users of multiple DL frameworks (Figure~\ref{fig:ngraph}), and requires much smaller code changes to non-HE code to invoke---potentially only one line (Appendix~\ref{sec:appendix}).

There have also been some recent compiler projects around HE. The {RAMPARTS} project~\cite{ramparts} translates models from the Julia language to HE operations implemented by {PALISADE} HE library~\cite{palisade}. No source code for the compiler is available, and source language support is limited to Julia. Moreover, the {PALISADE} library does not currently support CKKS, the HE scheme of choice for DL. The Cingulata~\cite{cingulata} compiler uses C++ as a source language and targets a custom implementation of the Fan-Vercauteren HE scheme. Because it first translates computations into boolean circuits rather than arithmetic compute graphs, it loses performance on DL operations such as GEMM. SHEEP~\cite{sheep} describes an abstract ISA for HE, and includes several HE schemes and implementations. However, the low-level language and lack of compiler make it difficult to use as a data science tool.

The CHET project~\cite{dathathri2018chet} also describes an ISA and a compiler for HE, but adopts a different approach from nGraph-HE. Whereas CHET performs compiler optimizations for HE at code generation time, just as any traditional compilation approach, nGraph-HE elevates optimizations for HE to the DL framework level, similarly to what frameworks such as TVM~\cite{tvm} do. Furthermore, to date, nGraph-HE is the only existing open-source framework for privacy-preserving DL. nGraph-HE's ability to support existing DL frameworks such as TensorFlow with minimal code changes is vital for data scientists, who must be able to rapidly prototype the accuracy and performance of DL models.

% Finally, the {CHET} project ~\cite{dathathri2018chet} also describes an ISA and runtime for HE, with a compiler that uses computation graphs of unspecified syntax as a source language, and no source code is provided. Finally, unlike nGraph-HE, no HE compiler project enables multiple DL frameworks such as TensorFlow, MXNet, or PyTorch by operating on their Python source code directly, and none exploits the rich DL compiler ecosystem that will continue to expand in support of further DL models, applications and hardware accelerators.

\subsection{The power of graph compilers}

DL frameworks, such as TensorFlow, MXNet and PyTorch, have greatly accelerated the development of DL models, allowing data scientists to express DL operations in high-level languages that can be easily ported from one platform to another (from a laptop to a
cloud-based server, {\it e.g.}). Graph compilers, such as the open-source Intel nGraph, have recently been developed to attack the challenge of optimizing performance on multiple DL frameworks and hardware targets. Compiling high-level framework code into an IR---a computation graph---removes the need to generate optimized code for each (framework, hardware target) pair. Instead, in order to use a new hardware target for all frameworks, one only needs to develop optimized code for each operation in the computation graph for the targeted hardware. In addition, the graph can be an advantageous representation for reasoning about higher-level optimizations, such as fusing operations, vectorization, etc.

These advantages apply directly to expressing DL computations using HE. By
treating HE schemes and the operations they support as instructions on
a virtual machine~\cite{helib}, we can enable HE computations
to a large set of DL frameworks while providing a clean separation
between DL and HE technologies. Moreover, as in the case of deep
learning, the graph representation facilitates various HE-specific
optimizations, such as reducing computation depth and identifying
opportunities for parallelism.

%% file: implementation.tex
\section{nGraph-HE}

We first describe the API adopted by nGraph-HE, as well as the mapping
onto two currently supported cryptosystems: BFV~\cite{bfv} and CKKS~\cite{ckks}, both implemented by the SEAL encryption library~\cite{seal}. We then discuss compile-time and runtime optimizations used in nGraph-HE. These include HE-specific optimizations that exploit the capabilities of the underlying cryptosystems, as well as parallelization methods to reduce execution time. Lastly, we discuss how to support additional cryptosystems.

 One difficulty in providing a unified framework for HE has been the variety of APIs and supported operations for various HE schemes. Following~\cite{HomomorphicEncryptionAPIStandard}, our API has {three components: (1) a \emph{cryptographic context}, which contains the static parameters of the encryption scheme, (2) a \emph{payload representation}, which contains the data, and (3) an \emph{assembly language}, which describes the functions implemented by the encryption scheme.

\subsection{Cryptographic context}
The cryptographic context stores the parameters of the cryptosystem, which consist of:
\begin{itemize}
	\item \emph{polynomial modulus degree} $(N)$;
	\item \emph{plaintext moduli} $(t = \prod_i t_i)$;
	\item \emph{ciphertext moduli} $(q = \prod_{i=1}^L q_i)$;	
	\item \emph{security level} $(\lambda$);
	\item \emph{HE scheme} as a unique string representation.
\end{itemize}
Depending on the cryptosystem, one or more of these parameters may not be required. The HE scheme implementations we currently support do not include bootstrapping; as such, we expect the cryptographic context to include enough ciphertext moduli to support the multiplicative depth of the DL model, i.e., $L \geq L_f$. The user will specify the cryptographic context as a command-line variable. In nGraph-HE, the ``HEBackend'' class stores the cryptographic context, as well as instantiations of (public, secret, relinearization) key tuples.
% \hl{ The cryptographic context is used to generate a public key and secret key, as well as relinearization keys.}

\subsection{Payload representation}

The payload representation stores the data and consists of plaintext and ciphertext representations. In nGraph-HE, the payload is stored in the ``HETensor'' class, which stores a pointer to an ``HEBackend'', necessary to obtain the keys. Figure~\ref{fig:notation} shows the relation between the terms in the payload representation.
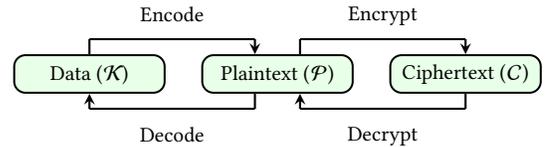
\begin{figure}[h!]
	
\tikzset{every picture/.style={line width=0.75pt}} %set default line width to 0.75pt
		
\tikzset{
	arrow/.style = {
		thick,->,>=stealth
	},
	arrowline/.style = {
		thick
	},
	object/.style ={
		rectangle, minimum width=2cm, minimum height=0.5cm, text centered, draw=black,rounded corners=.15cm,fill={rgb:green,1;white,10}
	}
}
\tikzset{font=\small}
\centering

\begin{tikzpicture}[node distance=2.5cm]
\def\dy{0.2cm}
\node (data) [object]{Data ($\mathcal{K}$)};
\node (plaintext) [object, right of=data]{Plaintext ($\mathcal{P}$)};
\node (ciphertext) [object, right of=plaintext]{Ciphertext ($\mathcal{C}$)};

%Encode arrow
\draw [arrowline] (data.north) -- ([yshift=\dy]data.north);
\draw [arrowline] ([yshift=\dy]data.north) -- ([yshift=\dy,xshift=-0.3cm]plaintext.north) node [midway, label=above:{Encode}] {};
\draw[arrow] ([yshift=\dy,xshift=-0.3cm]plaintext.north) -- ([xshift=-0.3cm]plaintext.north);

%Decode arrow
\draw [arrow] ([yshift=-\dy]data.south) -- (data.south);
\draw [arrowline] ([yshift=-\dy]data.south) -- ([yshift=-\dy,xshift=-0.3cm]plaintext.south) node [midway, label=below:{Decode}] {};
\draw [arrowline] ([yshift=-\dy,xshift=-0.3cm]plaintext.south) -- ([xshift=-0.3cm]plaintext.south);

%Encrypt arrow
\draw [arrowline] ([xshift=0.3cm]plaintext.north) -- ([yshift=\dy, xshift=0.3cm]plaintext.north);
\draw [arrowline] ([yshift=\dy, xshift=0.3cm]plaintext.north) -- ([yshift=\dy]ciphertext.north) node [midway, label={[yshift=-0.05cm]Encrypt}] {};
\draw[arrow] ([yshift=\dy]ciphertext.north) -- (ciphertext.north);

% Decrypt arrow
\draw [arrow] ([yshift=-\dy,xshift=0.3cm]plaintext.south) -- ([xshift=0.3cm]plaintext.south);
\draw [arrowline] ([yshift=-\dy,xshift=0.3cm]plaintext.south) -- ([yshift=-\dy]ciphertext.south) node [midway, label=below:{Decrypt}] {};
\draw [arrowline] ([yshift=-\dy]ciphertext.south) -- (ciphertext.south);
\end{tikzpicture}

	\caption{Relation between payload terms.}
	\label{fig:notation}
\end{figure}
Specifically, we have :
\begin{itemize}
\item \emph{data}: ($\mathcal{K}$). Usually $\mathcal{K}=\mathbb{R}^s$ or $\mathcal{K}=\mathbb{Z}^s$, $s \in \mathbb{N}$, a vector of numbers.
\item \emph{encode}: ($\mathcal{K}\rightarrow \mathcal{P}$). Uses the cryptographic context.
\item \emph{encrypt}: ($\mathcal{P}\rightarrow \mathcal{C}$). Uses the public key.
\item \emph{decrypt}: ($\mathcal{C}\rightarrow \mathcal{P}$). Uses the secret key.
\item \emph{decode}: ($\mathcal{P}\rightarrow \mathcal{K}$). Uses the cryptographic context.
\end{itemize}
This overall abstraction strictly generalizes the standard (encrypt, decrypt) model, since the encode and decode functions can be identity mappings. This allows us to store pre-computed plaintext values for optimization (Section~\ref{sec:special_plaintext_value_bpass}).

\subsection{Assembly language}
\label{sec:assembly}

We describe the assembly language of nGraph-HE in terms of nGraph operations~\cite{cyphers2018intel}. There are four low-level operations, which are typically supported by the APIs of HE cryptosystems:
\begin{itemize}
\item \texttt{Add}: $(\mathcal{C} \cup \mathcal{P}) \times \mathcal{C} \rightarrow \mathcal{C}$.
\item \texttt{Subtract}: $(\mathcal{C} \cup \mathcal{P}) \times \mathcal{C} \rightarrow \mathcal{C}$.
\item \texttt{Multiply}: $(\mathcal{C} \cup \mathcal{P}) \times \mathcal{C} \rightarrow \mathcal{C}$. For efficiency, the implementation should use relinearization if possible.
%\item \texttt{square}: $\mathcal{C} \rightarrow \mathcal{C}$. For some HE cryptosystems such as CKKS and BFV, square is an optimized multiply with the same operand. This operation may be optional depending on the cryptosystem.
\item \texttt{Negate}: $\mathcal{C} \rightarrow \mathcal{C}$.
\end{itemize}
Additionally, HE schemes will often implement a plaintext version ($\mathcal{P} \rightarrow \mathcal{P}$ or $\mathcal{P} \times \mathcal{P} \rightarrow \mathcal{P}$) of each operation.
Based on these low-level operations, nGraph-HE provides efficient parallelized implementations for the following compound operations: \texttt{AvgPool}, \texttt{Convolution}, and \texttt{Dot}. Developers can overwrite these default implementations with cryptosystem-specific optimizations.

nGraph-HE also provides implementations for the following tensor manipulation operations:
\texttt{Broadcast}, \texttt{Concat}, \texttt{Pad},  \texttt{Reshape}, \texttt{Reverse}, and \texttt{Slice}. See Table~\ref{tab:supported_ops} for a full list of supported operations and their mapping to TensorFlow operations.

\begin{table}[h!]
	\begin{center}
		\caption{Supported operations and mapping to TensorFlow operations.}
		\label{tab:supported_ops}
		\begin{tabular}{l l}
			\toprule
			\textbf{nGraph/nGraph-HE op} & \textbf{TensorFlow op} \\
			\midrule
			\texttt{Add} & \texttt{tf.add} \\
			\texttt{AvgPool} & \texttt{tf.nn.avg\_pool} \\
			\texttt{Broadcast} & \texttt{tf.broadcast\_to} \\
			\texttt{Concat} & \texttt{tf.concat} \\
			\texttt{Constant} & \texttt{tf.constant} \\
			\texttt{Convolution} & \texttt{tf.nn.convolution} \\
			\texttt{Dot} & \texttt{tf.matmul} \\
			\texttt{Multiply} & \texttt{tf.multiply, tf.square} \\
			\texttt{Negate} &\texttt{tf.negative} \\
			\texttt{Pad} & \texttt{tf.pad} \\
			\texttt{Reshape} & \texttt{tf.reshape} \\
			%			\texttt{Result} & \texttt{No TF equivalent} \\
			\texttt{Reverse} & \texttt{tf.reverse} \\
			\texttt{Slice} & \texttt{tf.slice} \\
			\texttt{Subtract} & \texttt{tf.subtract} \\
			\texttt{Sum} & \texttt{tf.reduce\_sum} \\
			\texttt{Parameter} & \texttt{tf.placeholder} \\
			\bottomrule
		\end{tabular}		
	\end{center}
\end{table}

Concretely, the nGraph-HE API consists of the following major components, shown in Figure~\ref{fig:architecture}:
\begin{itemize}
	\item \emph{Backend}. This stores the cryptographic context, and performs graph-level optimizations. The \emph{HESealBackend} and \emph{HEBFVBackend} classes inherit from \emph{HEBackend} class, which, in turn inherits from nGraph's \emph{Backend} class. 
	\item \emph{Tensor}. This stores the data. \emph{HEPlainTensor} and \emph{HECipherTensor} inherit from \emph{HETensor}, which in turn inherits from nGraph's \emph{Tensor} class. HEPlainTensors store \emph{HEPlaintext}'s, which is an abstract class from which \emph{seal::Plaintext} inherits; HECipherTensors operate analogously.
	\item \emph{Kernel}. The kernel consists of stand-alone implementations of nGraph \emph{op}s. Each implementation operates on \emph{HEPlaintext} and \emph{HECiphertext} inputs, which are dynamically cast to the appropriate cryptosystem-specific type at runtime. The \emph{Tensor} class hierarchy enables nGraph-HE to provide default implementations for each operation when no cryptosystem-specific implementation is present. This further decreases the overhead in adding a new cryptosystem to nGraph-HE.
\end{itemize}

\begin{figure}
	\tikzset{font=\small}	
\ifiacr
\begin{center}
	
	\tikzset{
		arrow/.style = {->,>=stealth},
		dotted_arrow/.style = {	->,>=stealth, dashed},
		doublearrow/.style ={<->, >=stealth	},
		arrowline/.style = {thick},
		ng_object/.style ={
			rectangle, minimum width=1.5cm, minimum height=0.5cm, text centered, draw=black,rounded corners=.15cm, fill={rgb:blue,1;white,10}
		},
		he_object/.style ={
			rectangle, minimum width=1.5cm, minimum height=0.5cm, text centered, draw=black,rounded corners=.15cm, fill={rgb:green,1;white,10}
		},
		thin_he_object/.style ={
			rectangle, minimum width=1.5cm, minimum height=0.5cm, text centered, draw=black,rounded corners=.15cm, fill={rgb:green,1;white,10}
		},
		seal_object/.style ={
			rectangle, minimum width=1.5cm, minimum height=0.5cm, text centered, draw=black,rounded corners=.15cm, fill={rgb:red,1;white,10}
		},
		thin_seal_object/.style ={
			rectangle, minimum width=1.5cm, minimum height=0.5cm, text centered, draw=black,rounded corners=.15cm, fill={rgb:red,1;white,10}
		},
		txt_object/.style ={
			rectangle, minimum width=1.5cm, minimum height=0.5cm, text centered, rounded corners=.15cm
		}
	}
\begin{tikzpicture}[node distance=2cm,x=1.3cm,y=0.7cm]
\else
\tikzset{
	arrow/.style = {
		->,>=stealth
	},
	dotted_arrow/.style = {
		->,>=stealth, dashed
	},
	doublearrow/.style ={
		<->, >=stealth
	},
	arrowline/.style = {
		thick
	},
	ng_object/.style ={
		rectangle, minimum width=1cm, minimum height=0.5cm, text centered, draw=black,rounded corners=.15cm, fill={rgb:blue,1;white,10}
	},
	he_object/.style ={
		rectangle, minimum width=1cm, minimum height=0.5cm, text centered, draw=black,rounded corners=.15cm, fill={rgb:green,1;white,10},text width=1.5cm
	},
	thin_he_object/.style ={
		rectangle, minimum width=1cm, minimum height=0.5cm, text centered, draw=black,rounded corners=.15cm, fill={rgb:green,1;white,10},text width=0.7cm
	},
	seal_object/.style ={
		rectangle, minimum width=1cm, minimum height=0.5cm, text centered, draw=black,rounded corners=.15cm, fill={rgb:red,1;white,10},text width=1.4cm
	},
	thin_seal_object/.style ={
		rectangle, minimum width=1cm, minimum height=0.5cm, text centered, draw=black,rounded corners=.15cm, fill={rgb:red,1;white,10},text width=1.0cm
	},
	txt_object/.style ={
		rectangle, minimum width=1cm, minimum height=0.5cm, text centered, ,rounded corners=.15cm
	}
}
\begin{tikzpicture}[node distance=2cm,x=0.9cm,y=0.7cm]
\fi
\def\dx{1}
\def\dy{1.5}

% Backend
\node(txt_backend) at (3*\dx,5.5*\dy) [txt_object]{\uline{Backend}};
\node(ng_backend) at (3*\dx,5*\dy) [ng_object]{ngraph::Backend};
\node(he_backend) at (3*\dx,4*\dy) [he_object]{HEBackend};
\ifiacr
\node(he_seal_backend) at (3*\dx,\dy) [seal_object]{HESealBackend};
\else
\node(he_seal_backend) at (3*\dx,\dy) [seal_object]{HESeal\\Backend};

\fi
% \node(he_seal_bfv_backend) at (4*\dx,0) [seal_object]{HESealBFV\\Backend};
% \node(he_seal_ckks_backend) at (2*\dx,0) [seal_object]{HESealCKKS\\Backend};

\draw [arrow] (ng_backend.south) -- (he_backend.north);
\draw [arrow] (he_backend.south) -- (he_seal_backend.north);
% \draw [arrow] (he_seal_backend.south) -- (he_seal_ckks_backend.north);
% \draw [arrow] (he_seal_backend.south) -- (he_seal_bfv_backend.north);

% Tensor
\node(txt_tensor) at (6*\dx, 5.5*\dy) [txt_object]{\uline{Tensor}};
\node(ng_tensor) at (6*\dx, 5*\dy) [ng_object]{ngraph::Tensor};
\node(he_tensor) at (6*\dx, 4*\dy) [he_object]{HETensor};

\ifiacr
\node(he_plain_tensor) at (5*\dx, 3*\dy) [he_object]{HEPlainTensor};
\node(he_cipher_tensor) at (7*\dx, 3*\dy) [he_object]{HECipherTensor};
\else
\node(he_plain_tensor) at (5*\dx, 3*\dy) [he_object]{HEPlain\\Tensor};
\node(he_cipher_tensor) at (7*\dx, 3*\dy) [he_object]{HECipher\\Tensor};
\fi
\node(he_plaintext) at (5*\dx, 2*\dy) [he_object]{HEPlaintext};
\node(he_ciphertext) at (7*\dx, 2*\dy) [he_object]{HECiphertext};

\ifiacr
\node(seal_plaintext) at (5*\dx, 1*\dy) [seal_object]{seal::Plaintext};
\node(seal_ciphertext) at (7*\dx, 1*\dy) [seal_object]{seal::Ciphertext};
\else
\node(seal_plaintext) at (5*\dx, 1*\dy) [seal_object]{seal::\\Plaintext};
\node(seal_ciphertext) at (7*\dx, 1*\dy) [seal_object]{seal::\\Ciphertext};
\fi

\draw [arrow] (ng_tensor.south) -- (he_tensor.north);
\draw [arrow] (he_tensor.south) -- (he_plain_tensor.north);
\draw [arrow] (he_tensor.south) -- (he_cipher_tensor.north);

\draw [arrow] (he_plain_tensor.south) -- (he_plaintext.north);
\draw [arrow] (he_cipher_tensor.south) -- (he_ciphertext.north);

\draw [arrow] (he_plaintext.south) -- (seal_plaintext.north);
\draw [arrow] (he_ciphertext.south) -- (seal_ciphertext.north);

% Kernel
\node(ng_op) at (9.5*\dx,5*\dy) [ng_object]{ngraph::op};
\node(add) at (8.7*\dx, 3*\dy) [thin_he_object]{Add};
\node(add_seal) at (8.7*\dx, 1*\dy) [thin_seal_object]{AddSeal};
\node(txt_tensor) at (9.5*\dx, 5.5*\dy) [txt_object]{\uline{Kernel}};
\node(conv) at (9.5*\dx, 4*\dy) [thin_he_object]{Conv};
\node(conv_seal) at (9.5*\dx, 2*\dy) [thin_seal_object]{ConvSeal};
\node(mult) at (10.3*\dx, 3*\dy) [thin_he_object]{Mult};
\node(mult_seal) at (10.3*\dx, 1*\dy) [thin_seal_object]{MultSeal};

\draw [arrow] (ng_op.south) -- (conv.north);

\draw [arrow] (ng_op.west) -| ([xshift=-0.2cm]add.north);
\draw [arrow] (ng_op.east) -| ([xshift=0.2cm]mult.north);

\draw [arrow] (conv.south) -- (add.north);
\draw [arrow] (conv.south) -- (mult.north);
\draw [dotted_arrow] (conv.south) -- (conv_seal.north);
\draw [arrow] (add.south) -- (add_seal.north);
\draw [arrow] (mult.south) -- (mult_seal.north);

\draw [arrow] (conv_seal.south) -- (add_seal.north);
\draw [arrow] (conv_seal.south) -- (mult_seal.north);
\end{tikzpicture}
\ifiacr
\end{center}
\else
\fi

\caption{Visualization of nGraph-HE architecture. Objects with the same color interact. The dotted line from Conv to ConvSeal indicates optional overriding of the default Conv implementation.}
\label{fig:architecture}
\end{figure}
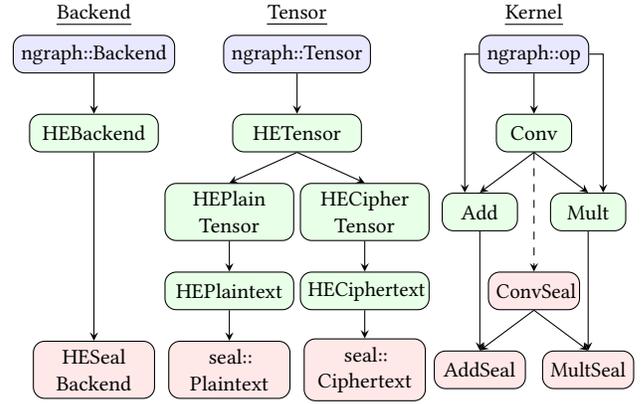

\subsection{Optimizations}
\label{sec:optim}
One of the benefits of using a compiler approach to homomorphic computation is the ability to perform optimizations that exploit the structure of the computation, the underlying cryptosystem, and the hardware. To illustrate this benefit, we implemented three classes of optimizations of which the first is run-time, and the second two are compile-time optimizations: (1) detection of special plaintext values; (2) mapping ISA-level parallelism in the privacy-preserving abstraction layer onto the parallel structures found in HE and modern microprocessors; and (3) graph-level optimizations.

\subsubsection{Special plaintext value bypass}
\label{sec:special_plaintext_value_bpass}

Operations between a ciphertext and a plaintext may arise when either the model or the data are encrypted, but not both. When performing such operations, nGraph-HE detects special values in the plaintext and, when possible, bypasses the corresponding HE operations. These runtime optimizations are HE-specific strength-reduction optimizations. Specifically, where $c \in \mathcal{C}$ is a ciphertext, and $p(i) \in \mathcal{P}$ is the plaintext encoding of $i$ we implement:
\begin{itemize}
\item $c \pm p(0)$: bypass HE operations and return $c$;
\item $c\ \times\ p(0)$: bypass HE operations and return a freshly-encrypted zero ciphertext, thereby resetting the noise budget;
\item $c \times p(1)$: bypass HE operations and return $c$;
\item $c \times p(-1)$: return the negation of $c$, avoiding an expensive multiply operation.
\end{itemize}
Bypassing HE operations not only reduces or resets encryption noise accumulation but also reduces runtime. One benefit of using a graph compiler is that higher-level compound operations, such as \texttt{Dot} and \texttt{Convolution}, automatically inherit the benefits of these optimizations. For instance, in a binarized neural network with binary convolution kernels, applying a \texttt{Convolution} operation will not invoke any calls to \texttt{Multiply}. To accommodate different binarization settings, we allow the user to independently enable or disable the Optimized Multiply and Optimized Addition plaintext value bypass. We demonstrate some of these runtime benefits quantitatively in Section~\ref{sec:gemm} and Section~\ref{sec:cryptonets}.

\subsubsection{Parallel operations:}
\label{sec:simd_packing}

\textbf{(1) HE-SIMD packing}. Some HE schemes (including BFV and CKKS) support Single Instruction Multiple Data (SIMD) operations~\cite{simd}, which we refer to as ``HE-SIMD'' operations to avoid confusion with the usage of the ``SIMD'' term in computer architecture. In simple terms, a vector of payload values can be encoded and encrypted as a single ciphertext, and operations on this ciphertext are equivalent to the same HE-SIMD operation on the values in the vector individually. nGraph-HE utilizes HE-SIMD packing across the mini-batch dimension, as in~\cite{gilad2016cryptonets}. Concretely, given a 4D tensor with shape $(N,C,H,W)$ format (batch size, channels, height, width), which typically requires $N \times C \times H \times W$ ciphertexts, nGraph-HE uses HE-SIMD packing to store the tensor as a 3D tensor of $C \times H \times W$ ciphertexts, with each ciphertext packing $N$ values. As shown in Section~\ref{sec:cryptonets}, running models with different mini-batch sizes (within the maximum allowed size) gives near-identical runtimes, significantly increasing the inference throughput. Analogously, loop unrolling in standard compiler optimization selects the amount of unrolling to maximize the number of utilized slots in vectorized operations. Here, increasing the batch size maximizes the number of utilized slots in each ciphertext.

Recent work~\cite{dathathri2018chet,juvekar2018gazelle} has experimented with using HE-SIMD packing to store NCHW-formatted tensors as different 2D or 3D tensors, improving on both memory usage and runtime on small batch sizes. However, although efficient operations for convolution and dot product exist in these packing schemes, reshape and broadcast operations become more complicated, and the ciphertext slots are more difficult to utilize entirely. Hence, our choice of HE-SIMD packing represents a performance tradeoff, optimizing for throughput and simplicity of implementation over latency.

\textbf{(2) OpenMP parallelization}. nGraph-HE makes extensive use of OpenMP~\cite{dagum1998openmp}, an API for shared-memory programming, for parallelization. It is used in data encryption and decryption, unary and binary element-wise operations, GEMM operations, and convolution. Different from HE-SIMD packing, OpenMP parallelization is applied to non-mini-batch dimensions. For instance, to encrypt a batch of 1024 images with shape $28\times28$, nGraph-HE encrypts the values at the first pixel location across all 1024 images as one ciphertext with HE-SIMD packing, and does so for all 784 pixel locations in parallel with OpenMP, resulting in a total of 784 ciphertexts. OpenMP parallelization significantly reduces the inference latency of our system.

\subsubsection{Graph-level optimizations}
One advantage of graph compilers is the ability to offer higher-level optimizations based on the computation graph. We briefly describe several graph optimizations analogous to standard compiler optimization of constant propagation and which are particularly relevant for HE.
\begin{itemize}
	\item \textit{AvgPool folding}. An AvgPool layer with window size ${s_1 \times s_2}$, followed by a Convolutional (Conv) layer with weights $W$ is replaced by the equivalent ScaledMeanPool operation followed by a Conv layer with weights $W / (s_1 \times s_2)$. This reduces the multiplicative depth $L_f$ from two to one.

	\item \textit{Activation folding}. A Conv or Fully Connected (FC) layer with weights $W$ followed by a polynomial activation of the form $ax^2 + bx + c$ is equivalent to the same Conv or FC layer with weights $aW$, followed by a polynomial activation of the form $x^2 + (b/a)x + (c/a)$. This reduces $L_f$ from two to one.

	\item \textit{Batch-Norm folding}. A Conv or FC layer followed by a Batch-Norm (BN) has the form:
	$$ z = W * x; \hspace{1cm} \hat{z} = \frac{z - \mu_z}{\sqrt{\sigma_z^2 + \epsilon}}; \hspace{1cm} z_{BN} = \gamma \hat{z} + \beta$$
where $\gamma, \beta, \mu_z, \sigma_z$ are all fixed during inference. A na\"ive implementation	would require a multiplicative depth $L_f=2$: one to compute $z$, and one to compute $z_{BN} = \hat{\gamma} z +\hat{\beta}$, where $\hat{\gamma} = \left( \frac{\gamma}{\sqrt{\sigma_z^2 + \epsilon}}\right)$ and $\hat{\beta} = \left(\beta - 
	\frac{\gamma \mu_Z}{\sqrt{\sigma_Z^2 + \epsilon}} \right)$ are constants at inference. However, we can equivalently compute $z_{BN} = (W \hat{\gamma}) * x + \hat{\beta}$ where $(W \hat{\gamma})$ is also constant at inference. This simplified representation has multiplicative depth $L_f=1$.
\end{itemize}
These optimizations are also possible in non-HE settings; for instance, BN folding is implemented in TensorFlow. However, the reduction in $L_f$ makes these optimizations especially useful in HE models. For instance, AvgPool folding is used in the CryptoNets model~\cite{gilad2016cryptonets}. See Section~\ref{sec:cifar10} and Section~\ref{sec:graph_optimizations} for examples of BN folding.

\subsection{Ciphertext-plaintext operations}
\label{sec:pt_ct_ops}
HE implementations of ciphertext-plaintext operations $\mathcal{C} \times \mathcal{P} \rightarrow \mathcal{C}$ are typically much faster than implementations of ciphertext-ciphertext operations $\mathcal{C} \times \mathcal{C} \rightarrow \mathcal{C}$.
To take advantage of this performance gain, nGraph-HE allows for three distinct computing paradigms, based on the privacy needs of the application.
\begin{itemize}
	\item \emph{Encrypted data, unencrypted model}. This use case occurs when private data are obtained from medical patients, while the model is kept on a remote server. This paradigm is the fastest, as it allows the most number of $\mathcal{C} \times \mathcal{P}$ operations.
	
	\item \emph{Encrypted model, unencrypted data}. This is the case when a company deploys a proprietary model to untrusted servers.
	\item \emph{Encrypted data, encrypted model}. Here, both the model and data are kept encrypted for most privacy, at the cost of the slowest runtime. This use case might occur when a company deploys a proprietary model to perform computations on sensitive data on untrusted hardware.
\end{itemize}
For debugging purposes, nGraph-HE also offers each operation in plaintext: $\mathcal{P} \times \mathcal{P} \rightarrow \mathcal{P}$ or $\mathcal{P} \rightarrow \mathcal{P}$.

\subsection{Adding a new cryptosystem}
Currently, nGraph-HE supports two cryptosystems, each implemented by the SEAL encryption library: BFV and CKKS. To support another cryptosystem, one simply needs to implement the storage model and the low-level operations in the assembly language instructions described above. Most HE cryptosystems already include similar APIs, so the implementation is usually straightforward. 
%For example, the \texttt{negate} operation maps to \texttt{Evaluator::negate} in both SEAL cryptosystems. The \texttt{multiply} operation, on the other hand, requires a different implementation for the CKKS and BFV implementations, due to the use of scaling in the CKKS cryptosystem.
As shown in Section~\ref{sec:assembly}, nGraph-HE provides default implementations for higher-level compound ops such as \texttt{Dot} and \texttt{Convolution}, which may be overridden with more efficient cryptosystem-specific implementations by the developer.

\subsection{DL Framework Integration}
A critical aspect of a new software library is the ease of adoption. nGraph-HE leverages Intel nGraph-TensorFlow~\cite{ngraph-tf} for seamless integration with the TensorFlow DL library~\cite{abadi2016tensorflow}. Modifying existing TensorFlow code to use nGraph-HE requires adding only a single line of code. See Appendix~\ref{sec:appendix} for a full example. This makes nGraph-HE extremely easy to use. To use models from other DL frameworks, such as MXNet, ONNX, and PyTorch, nGraph-HE users must first export the DL model to nGraph's serialized format.

nGraph-HE supports most operations commonly found in neural networks. Table~\ref{tab:supported_ops} shows the full list of operations currently supported by nGraph-HE, and the corresponding translation from TensorFlow operations. Notably absent is the support for MaxPool and ReLU operations. This is because HE supports only addition and multiplication, through which MaxPool and ReLU operations cannot be expressed.

%% file: results.tex
\section{Evaluation}

We tested nGraph-HE on a dual-socket Intel Xeon Platinum 8180 2.5GHz system with 376GB of RAM running Ubuntu 16.04. We used SEAL's implementation of CKKS and floating-point numbers for these measurements, although we have also tested nGraph-HE with SEAL's BFV implementation. We report two main findings. First, we leverage our compiler framework to implement HE-specific optimizations on small computation graphs. Second, we demonstrate the ease of implementing convolutional neural networks using a popular DL framework (TensorFlow), on both the MNIST~\cite{lecun1998mnist} and CIFAR-10~\cite{krizhevsky2014cifar} datasets. For the MNIST dataset, we additionally:
\begin{itemize}
	\item verify that the additional software layers through nGraph-HE to the underlying HE library impose minimal overhead;
	\item demonstrate the HE-SIMD packing (Section~\ref{sec:simd_packing}) and special plaintext value bypass (Section ~\ref{sec:special_plaintext_value_bpass}) optimizations;
	\item show the runtime dependence on three computing paradigms: encrypted data, encrypted model, encrypted data and model.
\end{itemize}
For the CIFAR-10 dataset, we also demonstrate the BN folding optimization.

\subsection{GEMM operations}
\label{sec:gemm}
\begin{figure}
\includegraphics[width=0.9\columnwidth]{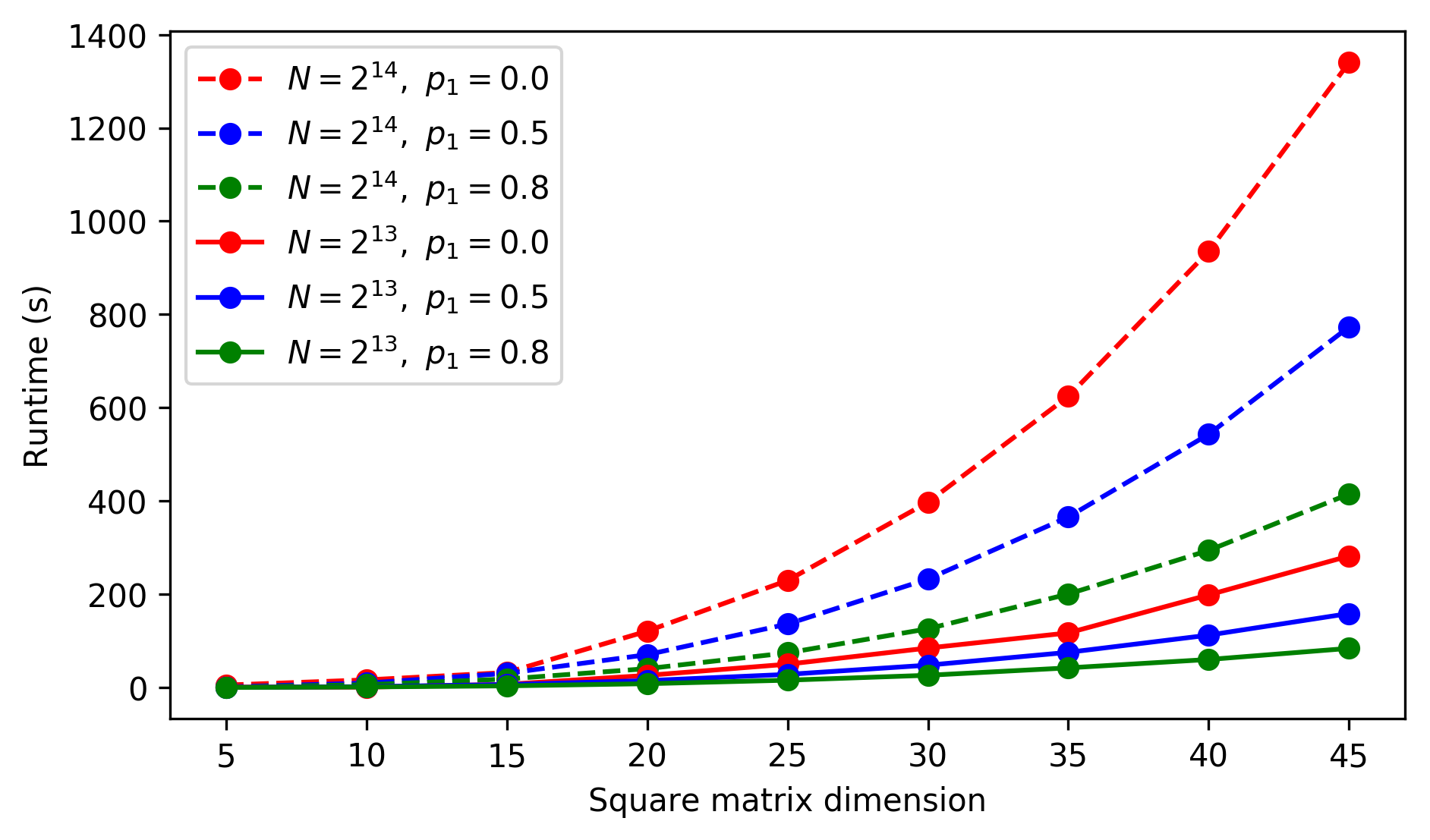}
\caption{Single-threaded runtime on GEMM operations as a function of matrix size, polynomial modulus, and sparsity.}
\label{fig:perf}
\end{figure}

We first tested nGraph-HE on general matrix-matrix multiplication (GEMM) operations, since these form the backbone of DL workloads. Figure~\ref{fig:perf} shows the runtime of computing $AB + C$, where
$A$, $B$, and $C$ are $n \times n$ matrices of random integers,
and where $A \in \mathcal{C}$, while $B, C \in \mathcal{P}$. (This
corresponds to the encrypted data, unencrypted model use case where, $A$ contains a user's
data while $B$ and $C$ are model weights.) To demonstrate two different parameter settings, we set the polynomial modulus degree to $N=2^{13}$ and $N=2^{14}$, and used SEAL's default ciphertext modulus for $\lambda=128$-bit security.

To illustrate the power of enabling HE using graph compilers, we perform the Optimized Multiply special plaintext value bypass on the $\mathcal{P} \times \mathcal{C}$ multiplication operation (Section~\ref{sec:special_plaintext_value_bpass}). We then measured the runtime savings by randomly setting 50\% and 80\% of the $B$ matrix to 1. These results correspond to the $p_1 = {0.5, 0.8}$ curves in Figure~\ref{fig:perf}. Because multiplication is more expensive than in most HE schemes, the runtime improvement is significant. The larger point, however, is that providing HE in the context of a graph compiler enables developers to provide HE-specific optimizations to the backend while data scientists continue to use the DL framework of their choice, treating HE as just another (virtual) hardware target.

\subsection{Graph-level optimizations}
\label{sec:graph_optimizations}
To demonstrate the utility of graph-level optimizations, we show an example of BN folding. We perform Convolution on a 3-channel $10 \times 10$ input of shape (1,3,10,10) using 4 kernels per channel, each of size $5 \times 5$, followed by BN. We consider two choices of parameters, each with security level $128 < \lambda < 192$: (1) $N=2^{14}$, 7 50-bit coefficient moduli; (2) $N=2^{13}$, 4 50-bit coefficient moduli.\footnote{A given set of encryption parameters achieves security level $\lambda$ if the coefficient modulus is smaller than SEAL's default coefficient choice at the same $(N, \lambda)$ pair~\cite{seal}. For instance $7 \times 50 = 350$, which is between SEAL's 305-bit $(N=2^{14}, \lambda=128)$ and 438-bit modulus $(N=2^{14}, \lambda=192)$, hence we achieve security level $128 < \lambda < 192.$} Table~\ref{tab:bn_folding} shows a moderate ${\sim}4\%$ decrease in runtime using BN folding, as expected. The larger point, however, is that this HE-specific graph-level optimization reduces the multiplicative depth $L_f$, enabling smaller encryption parameters, thereby greatly improving both runtime (${\sim}4$x) and memory usage. %(Section~\ref{sec:cifar10}).
\begin{table}[h!]
	\small
	\begin{center}
		\caption{Single-threaded runtimes on Conv-BN function when encrypting the data, using nGraph-HE directly. Runtimes are averaged across 10 trials.}
		\label{tab:bn_folding}
		\begin{tabular}{c c c
				S[separate-uncertainty,
				table-number-alignment = center,
				table-figures-integer = 3,
				table-figures-decimal = 2,
				table-figures-uncertainty = 1]
				S[separate-uncertainty,
				table-number-alignment = center,
				table-figures-integer = 3,
				table-figures-decimal = 2,
				table-figures-uncertainty = 1]
				S[separate-uncertainty,
				table-number-alignment = center,
				table-figures-integer = 3,
				table-figures-decimal = 2,
				table-figures-uncertainty = 1]
				}
			\toprule
			\multirow{2}{*}{\centering $\boldsymbol N$} & \multirowcell{2}{\textbf{BN} \\ \textbf{folding}} & \multirow{2}{*}{\centering $\boldsymbol L_f$} & \multicolumn{3}{c}{\textbf{Runtime (s)}} \\ \cmidrule{4-6}
			&& & {\textbf{Conv}} & {\textbf{BN}} & {\textbf{Total}} \\ \midrule
$2^{14}$ & \xmark & 2 & 130.83 \pm 1.14 & 6.28 \pm 0.12 & 137.24 \pm 1.21 \\ 
$2^{14}$ & \cmark & 1 & 130.57 \pm 1.57 & 0.25 \pm 0.01 & 130.97 \pm 1.57 \\
$2^{13}$ & \cmark & 1 & 33.06 \pm 0.68 & 0.06 \pm 0.0 & 33.16 \pm 0.68 \\
\bottomrule
	\end{tabular}
	\end{center}
\end{table}

\subsection{Neural networks}
\label{sec:cryptonets}
Next, to demonstrate the ease of using nGraph-HE, we implement neural networks on the standard MNIST dataset~\cite{lecun1998mnist}, as well as the more challenging CIFAR-10 dataset~\cite{krizhevsky2014cifar}.

\subsubsection{MNIST}
The MNIST dataset consists of handwritten digits, 60,000 for training, and 10,000 for testing, and is a standard benchmark for DL on HE. The original CryptoNets network~\cite{gilad2016cryptonets} is the standard HE-friendly network for MNIST, with architecture given in Appendix~\ref{sec:architecture}. Appendix~\ref{sec:appendix} shows the code to implement this network, which notably differs from the native TensorFlow code by just one line. We achieve an accuracy of ${\sim}$99\%, matching that reported in~\cite{gilad2016cryptonets}.

One concern with adding software abstractions is the runtime overhead. To measure this, we timed the network executing the TensorFlow code with nGraph-HE as the backend. This incurs the overhead of TensorFlow, the nGraph-TensorFlow bridge, and nGraph IR compilation. Within this execution, we separately time the sections that are also used in the execution of a C++ application that executes the serialized network natively.

Table~\ref{tab:mnist} shows the runtimes of these experiments, using ${N=2^{13}}$, ${N=2^{14}}$. We use SEAL's first 7 30-bit ciphertext coefficient moduli for CKKS (i.e., $q = \prod_{i=1}^7 q_i$, with each $q_i$ consisting of 30-bits), for security levels $128 < \lambda < 192$ and $\lambda > 256$, respectively. Note that the differences in times between the fourth and fifth columns (0.02s and 0.03s), which capture the overhead of graph compilation and bridging nGraph to TensorFlow, represent less than 0.1\% of overall runtime.
\begin{table}[!ht]
	\small
\begin{center}
	 \caption{Runtimes on CryptoNets network with and without the overhead of
 		TensorFlow integration and graph compilation.
 		Runtimes are averaged across 10 trials. Amortized runtimes are reported per image using batch size $N/2$ for maximum throughput and HE-SIMD slot utilization.}
 	\label{tab:mnist}
 	\ifiacr
 	\begin{tabular}{ c c c c c c }
 	\else
 	\begin{tabular}{ c c c p{1.7cm} p{2.0cm} p{1.3cm}}
 	\fi
\toprule
\multirow{2}{*}{$\boldsymbol N$} & \multirow{2}{*}{$\boldsymbol L_f$} & \multirowcell{2}{\textbf{Acc.} \\ \textbf{(\%)}} & \multicolumn{3}{c}{\textbf{Runtime (s)}} \\ \cmidrule{4-6}
& & & \centering \textbf{nGraph-HE} & \centering \textbf{TF+nGraph-HE} & \centering{\textbf{Amortized}}
\tabularnewline
\midrule
$2^{13}$ & 5 & \centering ${\sim}$99 & \centering{16.70 $\pm$ 0.23} & \centering{16.72 $\pm$ 0.23} & \centering{0.004} \tabularnewline
$2^{14}$ & 5 & \centering ${\sim}$99 & \centering{41.91 $\pm$ 1.58} & \centering{41.94 $\pm$ 1.58} & \centering{0.005} \tabularnewline
\bottomrule
\end{tabular}
\end{center}
\end{table}

\begin{figure}[!ht] 
	\begin{center}
		\includegraphics[width=\columnwidth]{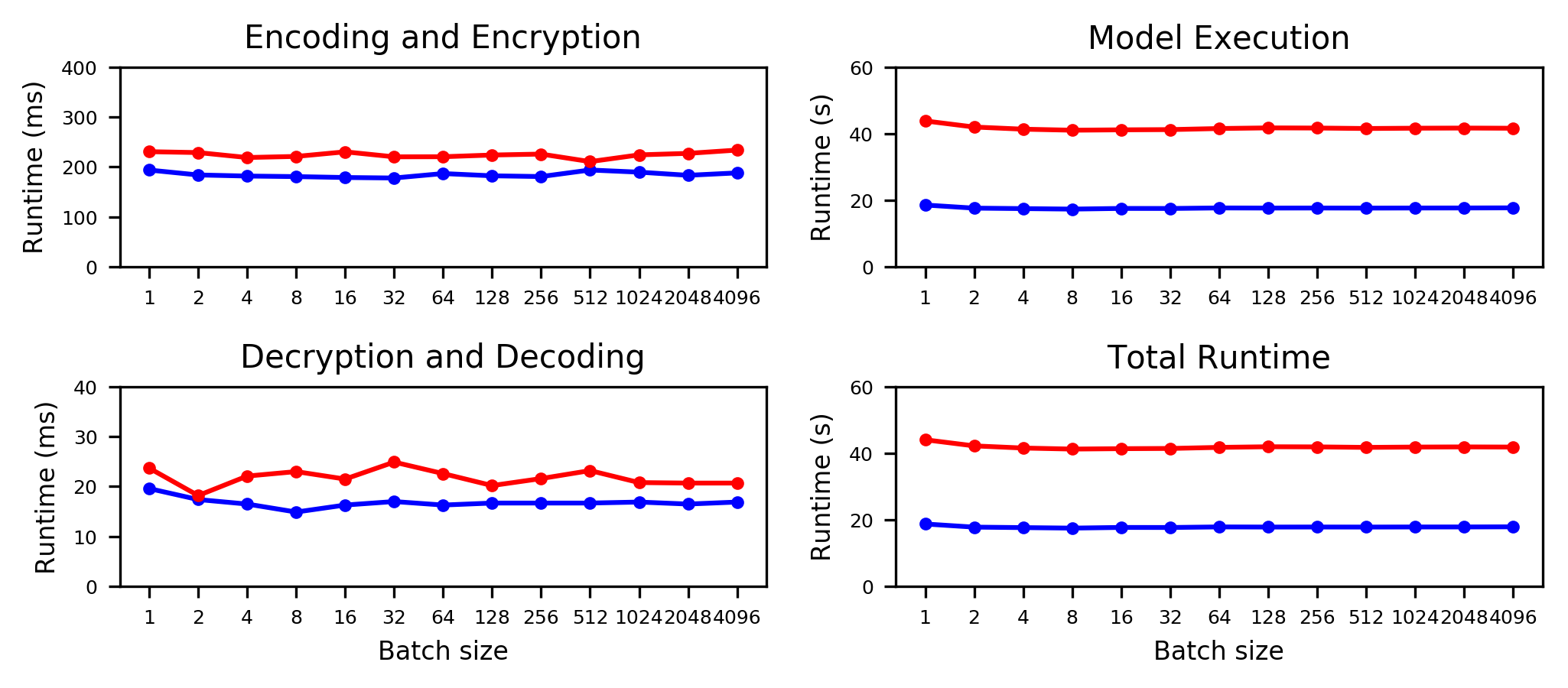}
		
		\newcommand{\redline}{\raisebox{2pt}{\tikz{\draw[-,red,solid,line width = 0.9pt](0,0) -- (5mm,0);}}}
		\newcommand{\blueline}{\raisebox{2pt}{\tikz{\draw[-,blue,solid,line width = 0.9pt](0,0) -- (5mm,0);}}}
		
		\caption{Runtimes on pre-compiled CryptoNets network with
			HE-SIMD packing for different batch sizes, for $N=2^{13}$ (\protect\blueline) and $N=2^{14}$ (\protect\redline).}
		\label{fig:simd}
	\end{center}
\end{figure}

Another benefit of using a graph compiler with HE is that the computation graphs provide opportunities to identify parallelism that can be exploited by some HE schemes, such as the ability to perform ``HE-SIMD'' operations on vectors of payload values (Section~\ref{sec:optim}). We implemented this capability and demonstrate it on the CryptoNets network. Figure~\ref{fig:simd} shows the CryptoNets inference runtime using batch sizes of 1 to 4096 for $N=2^{13}, 2^{14}$. We picked a maximum batch size of $2^{12}=4096$ because CKKS performs HE-SIMD operations on at most $N/2$ packed values. Note that runtimes are independent of batch size, for each step in running the network. Batching increases throughput significantly: for example, for the $N=2^{13}$ case, using a batch size of 4096 leads to an amortized runtime of 4.1ms per image, compared to 16.7s for a batch size of 1.

Another optimization nGraph-HE implements is $\mathcal{C} \times \mathcal{P}$ operations, which are typically much faster than $\mathcal{C} \times \mathcal{C}$ operations, and enable the three computing paradigms (encrypted data, model, or both) discussed in Section~\ref{sec:pt_ct_ops}. Using the CryptoNets network and same cryptographic context, Table~\ref{tab:encrypt_paradigm} shows the fastest runtime is achieved when just the data is encrypted. Encrypting the model incurs ${\sim}3.2$x runtime penalty, whereas encrypting the data and the model incurs ${\sim}3.6$x runtime penalty. Users can switch between the computing paradigms, enabling users to measure the privacy-performance tradeoff.

\begin{table}[h!]
	\begin{center}
		\caption{Runtimes on CryptoNets network when encrypting the data, the model, or both, using the TensorFlow nGraph-HE integration. Runtimes are in seconds and averaged across 10 trials.}
		\label{tab:encrypt_paradigm}
\begin{tabular}{c c
		S[separate-uncertainty,
		table-number-alignment = center,
		table-figures-integer = 3,
		table-figures-decimal = 1,
		table-figures-uncertainty = 1]
		S[separate-uncertainty,
		table-number-alignment = center,
		table-figures-integer = 3,
		table-figures-decimal = 1,
		table-figures-uncertainty = 1]}
	\toprule
	\multirow{2}{*}{$\boldsymbol N$} & \multicolumn{3}{c}{\textbf{Encrypt}} \\ \cmidrule{2-4}
	& {\textbf{Data}} & {\textbf{Model}} & {\textbf{Data and model}} \\
	\midrule
	$2^{13}$ & 16.7 $\pm$ 0.2 & 53.3 \pm 1.0 & 59.5 \pm 1.7 \\
	$2^{14}$ & 41.9 $\pm$ 1.6 & 128.2 \pm 1.2 & 142.6 \pm 9.9 \\
	\bottomrule
\end{tabular}		
	\end{center}
\end{table}
Finally, to demonstrate the benefit of special plaintext value bypass (Section~\ref{sec:special_plaintext_value_bpass}), we implement a binarized neural network. We adapt the CryptoNets network by binarizing each weight in the FC and Conv layers to $\{-1,1\}$, using the approach in~\cite{courbariaux2015binaryconnect}. To mitigate the diminished accuracy, we further add a non-binarized BN layer after each FC and Conv layer. BN-folding is disabled to preserve the binarization of the FC and Conv weights.

We consider two choices of parameters: 1) $N=2^{14}$ , 9 $30$-bit coefficient moduli, with security $192 < \lambda < 256$; 2) $N=2^{13}$, 7 $30$-bit coefficient moduli, with security $128 < \lambda < 192$. As shown in Table~\ref{tab:BNN}, enabling the Optimized Multiply special plaintext value bypass provides a moderate ${\sim}1.2$x speedup. However, a more significant ${\sim}3.7$x runtime speedup arises due to the lower multiplicative depth $L_f$, enabling a smaller choice of $N$. The runtime of 14.8s is faster than 16.7s in the original CryptoNets network, at the cost of reduced accuracy.

\begin{table}[h!]
	\begin{center}
		\caption{Runtimes on binarized CryptoNets network when encrypting the data, using the TensorFlow nGraph-HE integration. Runtimes are averaged across 10 trials.}
		\label{tab:BNN}
		\begin{tabular}{ c c c c c c c}
			\toprule
			$\boldsymbol N$ & $\boldsymbol L$ & $\boldsymbol L_f$ & \textbf{Optimized Mult.} & \textbf{Runtime (s)} & \textbf{Acc. (\%)} \\ \midrule
			$2^{14}$ & 9 & 8 & \xmark & \centering $55.2 \pm 2.3$ & \textbf{$96.9 \pm 0.5$} \\
			$2^{14}$ & 9 & 5 & \cmark & $45.2 \pm 1.3$ &\textbf{$96.9 \pm 0.5$} \\
			$2^{13}$ & 7 & 5 & \cmark & $14.8 \pm 0.9$ & \textbf{$96.9 \pm 0.5$} \\ \bottomrule
		\end{tabular}		
	\end{center}
\end{table}

\subsubsection{CIFAR-10}
\label{sec:cifar10}
The CIFAR-10 dataset is a standard image classification dataset consisting of 60,000 color images of shape $32 \times 32 \times 3$, of which 50,000 are used for training, and 10,000 are used for testing, and with 6,000 examples for each of 10 different classes. The larger image size and color channels make CIFAR-10 a significantly more challenging task than MNIST. There is currently no seminal CIFAR-10 HE-friendly network as there is for MNIST.
%Although~\cite{hesamifard2017cryptodl} and~\cite{dathathri2018chet} implement CIFAR-10 HE-friendly networks, these deeper models are more difficult to train.
Due to the use of unbounded polynomial activations, numerical overflow during training is prevalent, although gradient clipping and BN somewhat mitigate this effect. We implement a CIFAR-10 network with architecture given in Appendix~\ref{sec:architecture}. To demonstrate the versatility of our framework, we train the CIFAR-10 network in four different configurations, by toggling two independent settings:
\begin{itemize}
	\item \emph{BN}. If enabled, a BN layer is added after each Conv layer. During training, we use batch size $n=128$; during inference, we use HE-SIMD packing to enable batch size $n=8192$.
	\item \emph{Trained activations}. If enabled, each polynomial activation is of the form $ax^2 + bx$, with $a$ initialized to 0, and $b$ initialized to 1, and with $a,b$ updated during training. If disabled, each polynomial activation is $0.125x^2 + 0.5x + 0.25$, following the approach in~\cite{chou2018faster}.
\end{itemize}
Furthermore, to prevent numerical overflow during training, we clip the gradients to $[-0.25, 0.25]$. To demonstrate the advantage of our graph-level optimizations, we toggle the BN folding optimization where BN is used. The CIFAR-10 network has a multiplicative depth $L_f=8$, which is significantly deeper than the CryptoNets network, with $L_f=5$. In order to accommodate this additional depth, we choose 10 30-bit ciphertext coefficient moduli, and ${N=2^{14}}$ for security level $\lambda=192$ ~\cite{seal}. When BN folding is disabled, $L_f$ increases to 10. Accordingly, we use 11 30-bit ciphertext moduli for a reduced security level of $128 < \lambda < 192$.

\begin{table}[h!]
	\begin{center}
		\small
		\caption{Runtimes on CIFAR-10 network when encrypting the data, using the direct nGraph-HE integration. Runtimes and accuracies are averaged across 10 trials. Amortized runtimes are per image, using batch size $N/2$ for maximum throughput and HE-SIMD slot utilization.}
		\label{tab:cifar10}
		\ifiacr
			\begin{tabular}{ c c c c c c c}
		\else
			\begin{tabular}{ c p{0.3cm} c c c p{1.25cm} p{1.3cm}}
		\fi

			\toprule
			 \multirow{2}{*}{$\boldsymbol L$} & \multirow{2}{*}{\centering{$\boldsymbol L_f$}} & \multirow{2}{*}{\textbf{BN/fold}} & \multirow{2}{*}{\centering{\textbf{Act.}}} & \multirowcell{2}{\textbf{Accuracy} \\ \textbf{(\%)}} & \multicolumn{2}{c}{\textbf{Runtime (s)}} \\ \cmidrule{6-7}
			 % \centering{Runtime (s)} &\centering{Amortized Runtime} \tabularnewline
& & & & & \centering{\textbf{Total}} & {\textbf{Amortized}} \\ \midrule			 
$11$ & \centering{10} & \centering{\cmark}/ \centering{\xmark} & \centering{Train} & \centering 62.1 $\pm$ 6.4 &\centering{ 1628 $\pm$ 37 } & \centering{ 0.199} \tabularnewline
$11$ & \centering{8} & \centering{\cmark}/ \centering{\cmark} & \centering{Train} & \centering 62.1 $\pm$ 6.4 & \centering 1637 $\pm$ 42 & \centering{0.200} \tabularnewline
$10$ & \centering{8} & \centering{\cmark}/ \centering{\cmark} & \centering{Train} & \centering 62.1 $\pm$ 6.4 & \centering 1350 $\pm$ 22 & \centering{0.165} \tabularnewline \midrule
$11$ & \centering{10} & \centering{\cmark}/ \centering{\xmark} & \centering{Fix} & \centering 62.2 $\pm$ 3.5 & \centering{ 1641 $\pm$ 32 } & \centering{ 0.200} \tabularnewline
$11$ & \centering{8} & \centering{\cmark}/ \centering{\cmark} & \centering{Fix} & \centering 62.2 $\pm$ 3.5 & \centering 1651 $\pm$ 33 & \centering{0.202} \tabularnewline
$10$ & \centering{8} & \centering{\cmark}/ \centering{\cmark} & \centering{Fix} & \centering 62.2 $\pm$ 3.5 & \centering 1359 $\pm$ 19 & \centering{0.166} \tabularnewline \midrule
$10$ & \centering{8} & \centering{\xmark} & \centering{Tr} & \centering 55.6 $\pm$ 6.7 & \centering 1321 $\pm$ 20 & \centering{0.161} \tabularnewline \midrule
$10$ & \centering{8} & \centering{\xmark} & \centering{Fix} &\centering 57.8 $\pm$ 1.3 & \centering 1324 $\pm$ 13 & \centering{0.161} \tabularnewline \bottomrule
		\end{tabular}
	\end{center}
\end{table}

Table~\ref{tab:cifar10} shows the runtimes of the CIFAR-10 network, which are significantly higher than the MNIST CryptoNets network, due to the increased complexity of the model and dataset. BN provides a significant increase in accuracy, as polynomial activations are constrained within a narrower range. We observe $L_f$ is constant when BN folding optimization is enabled. Enabling BN-folding reduces $L_f$ from 10 to 8, with negligible speedup. However, the reduced multiplicative depth allows for use of fewer ciphertext moduli, which provides a ${\sim}1.2$x speedup.

%% file: extensions.tex
\section{Conclusion and Future Work}
\label{sec:extensions}
We have presented nGraph-HE, a backend to the Intel nGraph DL compiler, which enables DL on homomorphically encrypted data. nGraph-HE supports a variety of DL frameworks such as TensorFlow to allow easy adoption by data scientists. We have demonstrated the capability of nGraph-HE to implement networks on MNIST and CIFAR-10 with minimal computational and code overhead. Furthermore, we demonstrated several optimizations, including special plaintext value bypass, HE-SIMD packing, graph-level optimizations, and plaintext operations. The data scientist can take advantage of these optimizations with only minimal changes to their code, enabling rapid prototyping and development.

Looking ahead, an additional benefit of using graph compilers in the context of HE is the ability to extract the computational (especially multiplicative) depth of the computation, which is needed to set the security parameters of the HE scheme. A useful
extension of this work, therefore, would be to enable automatic
selection of HE parameters at compile time as a function of desired
security level. Another area for future work is to incorporate recent optimizations
for matrix operations in HE~\cite{juvekar2018gazelle}. Finally, we
would like to extend this framework so that it can also include hybrid
schemes that combine HE with multi-party-computation (MPC), such as garbled circuits~\cite{bellare2013efficient, songhori2015tinygarble}, or oblivious transfer~\cite{henecka2010tasty, demmler2015aby, buscher2018hycc}. Such hybrids have been shown~\cite{juvekar2018gazelle} to deliver much
faster performance at the expense of higher communication
costs. The optimal decomposition of a DL workload into HE
and MPC stages could be determined at compile time and would be
greatly facilitated by access to the underlying computation graph.

%% file: appendix.tex
\section{Appendix}

\subsection{Network Architectures}
\label{sec:architecture}
For each architecture, $n$ indicates the batch size.
\begin{itemize}
	\item 
CryptoNets, with activation $Act(x) = x^2$.
\begin{enumerate}
	\item \emph{Conv}. [Input: $n \times 28 \times 28$; stride: 2; window: $5 \times 5$; filters: 5, output: $n \times 845$] + \emph{Act}.
	\item \emph{FC}. [Input: $n \times 845$; output: $n \times 100$] + \emph{ Act}.
	\item \emph{FC}. [Input: $n \times 100$; output: $n \times 10$].
\end{enumerate} %\footnote{As in~\cite{gilad2016cryptonets}, for inference we omit the final softmax layer and squashed layers 3-6 into one linear layer, which we call layer (3).}:

\item 
Binarized CryptoNets, with activation $Act(x) = x^2$.
\begin{enumerate}
	\item \emph{BinaryConv}. [Input: $n \times 28 \times 28$; stride: 2; window: $5 \times 5$; filters: 5, output: $n \times 845$] + \emph{BN} + \emph{Act}.
	\item \emph{BinaryFC}. [Input: $n \times 845$; output: $n \times 100$] + \emph{BN} + \emph{Act}.
	\item \emph{BinaryFC}. [Input: $n \times 100$; output: $n \times 10$] + \emph{BN}.
\end{enumerate}

\item 
CIFAR-10 network, with polynomial activation.
\begin{enumerate}
	\item \emph{Conv}. [Input: $n \times 32 \times 32 \times 3$; stride: 2; window: $5 \times 5$; filters: 40, output: $n \times 40 \times 16 \times 16$] + (\emph{BN}) + \emph{Act}.
	\item \emph{AvgPool}. [Input: $n \times 32 \times 32 \times 3$; stride: 2; window: $5 \times 5$, filters: 40, output: $n \times 40 \times 8 \times 8$].
	\item \emph{Conv}. [Input: $n \times 40 \times 8 \times 8$; stride: 1; window: $3 \times 3$; filters: 80, output: $n \times 80 \times 8 \times 8$] +  (\emph{BN}) + \emph{Act}.
	\item \emph{FC}. [Input: $n \times 5120$, output: $n \times 10$].
\end{enumerate}
\end{itemize}

\subsection{CryptoNets Inference}
\label{sec:appendix}
Figure~\ref{fig:mnist_example} shows a full example of performing inference on the CryptoNets model. Note that the only modification required to enable nGraph-HE is the addition of the \texttt{import ngraph\_bridge} line.

\ifiacr
\else
\begin{center}
	\begin{figure*}[hb!]
		\begin{subfigure}[b]{0.6\textwidth}
\begin{lstlisting}[language=Python]
"""CryptoNets MNIST classifier"""
import ngraph_bridge  # <-- enable nGraph-HE
import numpy as np
from tensorflow.examples.tutorials.mnist import input_data
import tensorflow as tf

batch_size = 4096

mnist = input_data.read_data_sets(
    '/tmp/tensorflow/mnist/input_data', one_hot=True)

# Create inference network
parameter_0 = tf.placeholder(tf.float32, [None, 784])
reshape_5_7 = tf.reshape(parameter_0, [-1, 28, 28, 1])
constant_4 = tf.constant(
    np.loadtxt('W_conv1.txt', dtype='f').reshape([5, 5, 1, 5]))
convolution_8 = tf.nn.conv2d(
    reshape_5_7, constant_4, strides=[1, 2, 2, 1], padding='VALID')
multiply_10 = tf.square(convolution_8)
constant_3 = tf.constant([[0, 0], [0, 1], [0, 1], [0, 0]])
pad_11 = tf.pad(multiply_10, constant_3)
reshape_12 = tf.reshape(pad_11, [-1, 845])
constant_2 = tf.constant(
    np.loadtxt("W_squash.txt", dtype='f').reshape([845, 100]))
dot_13 = tf.matmul(reshape_12, constant_2)
multiply_14 = tf.square(dot_13)
constant_1 = tf.constant(np.loadtxt('W_fc2.txt', dtype='f').reshape([100, 10]))
y_conv = tf.matmul(multiply_14, constant_1)

# Run network
with tf.Session() as sess:
    x_test = mnist.test.images[:batch_size]
    y_conv_val = y_conv.eval(feed_dict={parameter_0: x_test})
\end{lstlisting}
			\caption{Python code to execute a trained CryptoNets model using TensorFlow}
		\end{subfigure}%
		\begin{subfigure}[b]{0.4\textwidth}
			\includegraphics[width=0.65\textwidth]{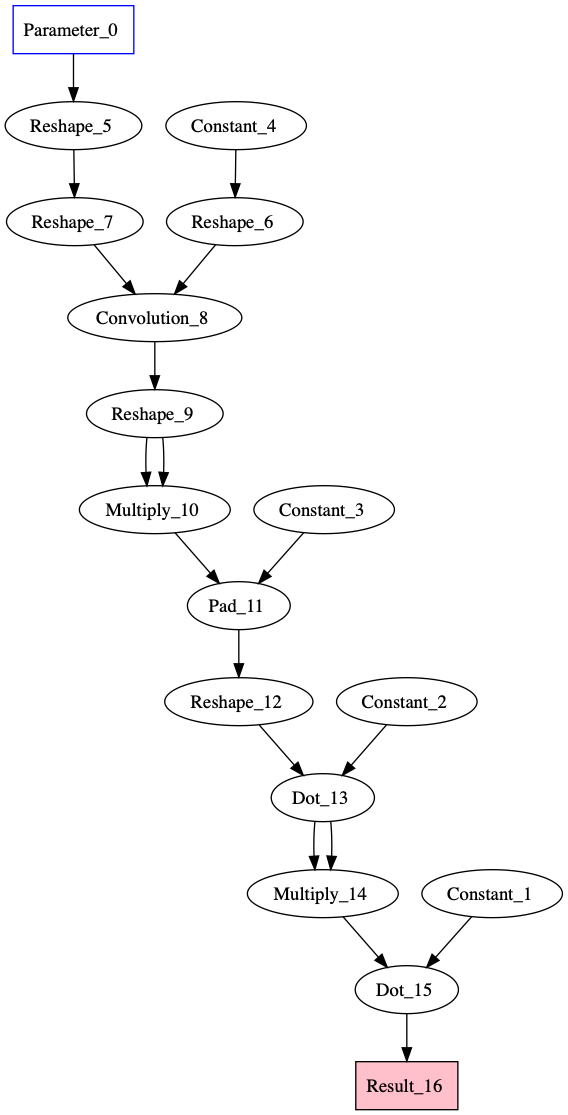}		
			\caption{Computational graph generated from the Python code by the nGraph compiler}
		\end{subfigure}%
		\caption{Source code and intermediate representation of the MNIST CryptoNets network.
			% \vspace{1.5cm}
		}
		\label{fig:mnist_example}
	\end{figure*}
\end{center}
\fi

%% file: artifacts.tex
\section{nGraph-HE Artifact Appendix}
Code and runtime artifacts to replicate runtime results are publicly available at \url{https://github.com/NervanaSystems/he-transformer/tree/v0.2-benchmarks-2}

Specifically, the \emph{benchmarks} folder contains detailed instructions on how to replicate the results, including our own runtime results from which the tables and figures were created.

Performance analysis completed on Jan 16 - Mar 21, 2019 using a Xeon Platinum 8180 platform with 112 CPUs operating at 2.5Ghz, 2 sockets, and 376GB of RAM running HE Transformer (v0.2) with nGraph-tf (v0.9.0) and nGraph (v0.11.0) on Ubuntu 16.04.4 LTS.